\documentstyle[sprocl,epsfig]{article}

\def\be{\begin{equation}}
\def\ee{\end{equation}}
\def\bea{\begin{eqnarray}}
\def\eea{\end{eqnarray}}

\begin{document}
\title{QCD POTENTIOLOGY}

\author{G. S. BALI
\footnote{Invited talk presented at ``4th International
Conference on Quark Confinement
and the Hadron Spectrum'', Vienna, 3--8 July 2000.}
}

\address{Department of Physics \& Astronomy, University of Glasgow,
Glasgow,\\ G12 8QQ, Scotland\\ E-mail: g.bali@physics.gla.ac.uk}

\maketitle\abstracts{We review the connection between
QCD and potential models of quarkonia in the framework
of effective field theories, with an emphasis on
non-perturbative methods such as lattice simulations.
The static and heavy quark potentials are introduced
and the applicability of the
non-relativistic and adiabatic approximations are
discussed. The r\^oles of so-called hybrid potentials
are explored and we comment on the QCD analogue of the
QED Lamb shift.}
\section{Potential models}
Long before direct non-perturbative
QCD evaluations became feasible potential models,
some QCD inspired and some not, turned out
to be extremely useful for an understanding of the masses
and decay rates of bottomonium ($\Upsilon$), charmonium ($J/\psi$)
and $B_c$ states. The underlying idea is that when the quark
mass $m$ is bigger than all other
bound state energy scales,
time scales $\Delta t_Q$ that characterise
the relative heavy quark movement become much bigger than
time scales $\Delta t_g$, associated with
the gluon (and sea quark) dynamics, too.
In this situation the leading Born-Oppenheimer (or adiabatic)
approximation should be applicable
and feedback effects of the moving
heavy quarks onto the accompanying
gluons and sea quarks might be neglected.
If in addition the typical relative quark velocity
is much smaller than the speed of light, $v=p/m\ll 1$, the bound state
should be governed by a Schr\"odinger Hamiltonian,
\begin{equation}
\label{eq:schroe}
H=2m+\frac{p^2}{2\mu_R}+V_{\mbox{\scriptsize phen}}(r),\quad
H\psi_{n}=E_n\psi_{n},
\end{equation}
where $\mu_R=m/2$ denotes the reduced mass
and the potential
$V_{\mbox{\scriptsize phen}}(r)$ parameterises the interaction that
is mediated by gluons and sea quarks.

While in lattice QCD calculations spectrum and wave functions can in
principle be predicted for a given set of quark masses and scale parameter,
in the potential model approach one is initially confronted
with the inverse problem of guessing or deriving the potential
$V_{\mbox{\scriptsize phen}}$ from the knowledge of some energy levels and decay rates,
before any predictions can be made. The experimental spin averaged
quarkonia levels seem to constrain the potential (up to an additive
constant) reasonably well
for distances $0.2$~fm~$<r<1$~fm to a form that
can be represented in terms of a funnel (or Cornell)
parametrisation,~\cite{Eichten:1975af}
\begin{equation}
\label{eq:funnel}
V_{\mbox{\scriptsize phen}}(r)=\sigma\,r-\frac{e}{r},
\end{equation}
with $e\approx 0.5$ and a string tension $\sigma\approx 425^2\mbox{~MeV}^2$.
The advantage of potential models over other approaches such as Lattice
Gauge Theory or Spectral Sum Rules is the relative ease with which properties
of radial excitations can be derived and the simplicity and transparency
of the representation of the bound state problem in terms of
a nonrelativistic quantum mechanical Hamiltonian.

If a potential model has anything to do with QCD, the
{\em heavy quark potential}
$V_{\mbox{\scriptsize phen}}(r)$ should
be calculable on the lattice and such a determination should
increase its predictive power. One object that is
easily accessible in lattice studies is the so-called
{\em static potential},
$V_0(r)$.
In this article we shall address the following questions:
does the ``heavy quark potential'' $V_{\mbox{\scriptsize phen}}$ exist?
What is the relation between $V_{\mbox{\scriptsize phen}}$ and $V_0$?
Can relativistic corrections systematically be incorporated into
Eq.~(\ref{eq:schroe})?
\section{QCD static potentials}
Ironically, forgetting everything one knows about quarks and their
properties like spin and mass seems to be a good strategy for
an understanding of heavy quark bound state problems.
After successfully mastering this first mental barrier
one has to place an external colour charge $Q$ and anti-charge
$Q^*$ at a distance $r$ into the QCD vacuum.
The static potential $V_0(r)$ is then the energy difference of
this configuration with respect to the vacuum energy.
Since we are dealing with a quantum field theory,
an additive mass renormalisation is required,
due to the self energy of the point like sources,
\begin{equation}
\label{eq:vself}
V_{0,\mbox{\scriptsize phys}}(r)=V_0(\mu;r)-V_{\mbox{\scriptsize self}}(\mu),\quad
V_{\mbox{\scriptsize self}}(\mu)\propto \mu/\ln\mu\quad(\mu\rightarrow\infty):
\end{equation}
the static potential
$V_0(\mu;r)$ diverges as the cut-off on gluon momenta $\mu$
is sent to infinity while, once a subtraction
scheme is specified, $V_{0,\mbox{\scriptsize phys}}$
is well defined.

$V_0(\mu;r)$ can be computed from Wilson loops,
for instance
on a lattice with spacing $a\simeq \pi/\mu$.
From perturbative QCD one would expect the generic
Coulomb short distance form,
$V_0(r)=-e/r$,
where the effective
$e=(4/3)\alpha_R(r)$ will logarithmically depend on the
distance $r$ while --- neglecting 
sea quarks --- one might expect
a linear rise at large distances, in agreement with
the funnel parametrisation, Eq.~(\ref{eq:funnel}).

In QCD the potential is the energy of a bound state of gluons and
static sources.
This contrasts the
situation in QED where confinement is lacking and
photons carrying any surplus energy will just be radiated away.
In QCD a discrete spectrum of
gluonic excitations exists, the so-called hybrid
potentials, see e.g.~\cite{Michael:2000jd,Bali:2000gf}.
This leads us to conclude that in addition to the distance $r$
another scale, $\Delta V(r)$, the gluonic binding energy,
has to be considered. In a quenched setting
$\Delta V(r=0)$
will be a glueball mass, the lightest of which is
about
1.6~GeV, while in the case of QCD with sea quarks, radiation of a pair of
pions becomes possible. While the coupling constant
$\alpha_s(r^{-1})\rightarrow 0$ as $r\rightarrow 0$,
$\alpha_s[\Delta V(r)]$ remains large at all
distances, from which one might expect perturbation
theory to break down when na\"{\i}vely applied to the Wilson loop.

\begin{figure}[t]
\centerline{\epsfig{figure=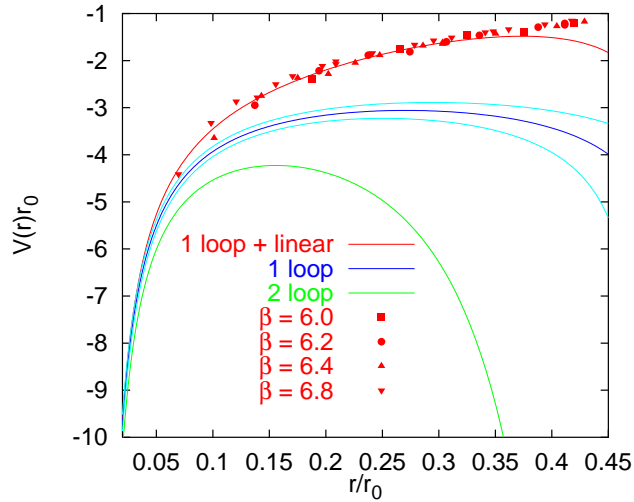,height=7.cm}}\vskip -.5cm
\caption{Comparison between the quenched lattice QCD potential
(full symbols) and perturbation
theory.~\protect\cite{Bali:2000gf,Bali:1999ai}\label{fig:potpert}}\vskip -.5cm
\end{figure}

In Fig.~\ref{fig:potpert} we compare the quenched QCD potential, 
calculated on the lattice with continuum perturbation
theory~\cite{Bali:1999ai,Bali:2000gf}
in units $r_0\approx 0.5$~fm at short distances, $r<0.23$~fm.
Note that the $\Lambda$-parameter has been determined from
an unrelated lattice observable, such that the only freedom
in this representation is an additive constant, due to different
values of $V_{\mbox{\scriptsize self}}$ in different regularisation schemes.
Indeed, the perturbative series does not seem to
convincingly approach the non-perturbative result and
the coefficients of the perturbative expansion are large
as indicated by the big difference between one- (order $\alpha_s^2$)
and two-loop (order $\alpha_s^3$)
potentials.
This means that different resummation prescriptions will in
general yield different two-loop results, some being closer to
the non-perturbatively determined potential, some disagreeing even more.

\begin{figure}[t]
\centerline{\epsfig{figure=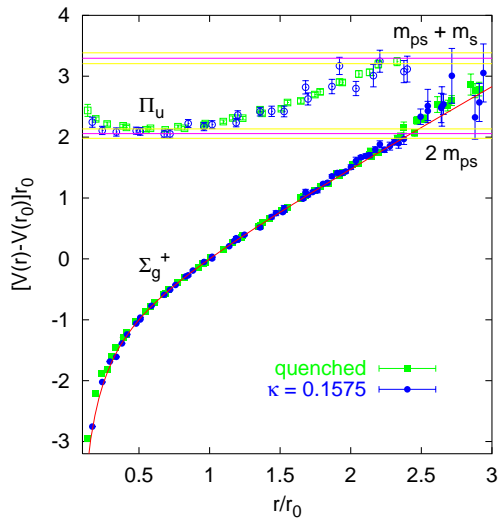,height=7.cm}}\vskip -.3cm
\caption{Static QCD potential and first hybrid excitation,
without (quenched) and with
($\kappa=0.1575$) sea
quarks.~\protect\cite{Bali:2000vr}\label{fig:potfull}}\vskip -.5cm
\end{figure}

Gluonic excitations can be classified with respect to representations of the
cylindrical symmetry group $D_{\infty h}$.
The lowest hybrid
excitation, in which the glue carries one unit of angular momentum
about the intersource axis, is labelled by $\Pi_u$.
A recent lattice result~\cite{Bali:2000vr}
of both the static potential ($\Sigma_g^+$) and
the hybrid potential
$\Pi_u$, normalised
such that $V_0(r_0)=0$
is depicted in Fig.~\ref{fig:potfull}, both without sea quarks (quenched) and
with 2 flavours of sea quarks, slightly lighter than the strange quark
($\kappa=0.1575$).

When including sea quarks the potential will eventually
flatten around~\cite{Bali:2000vr}
$r\approx 2.3\, r_0$ (string breaking)
while in the quenched case the linear rise will continue.
Within the region $r>0.2$~fm the lattice potentials can be fitted
to the parametrisation Eq.~(\ref{eq:funnel}), with the result
$e\approx 0.3$ neglecting sea quarks and $e\approx 0.36$ with two flavours
of light sea quarks.
\section{The infra red problem of perturbation theory}
In QED without fermions, the perturbative expansion of a Wilson loop
can schematically be written as,
\begin{equation}
\label{eq:pert}
\langle W(r,\tau)\rangle
=1+\mbox{\begin{minipage}{1.05cm}
\epsfig{file=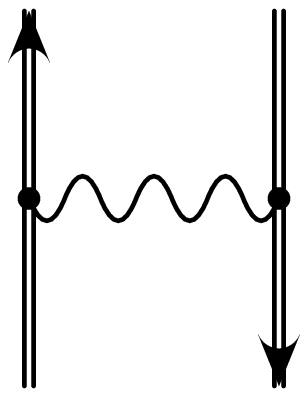,width=1cm}\end{minipage}}
+
\frac{1}{2}\,\mbox{\begin{minipage}{1.05cm}
\epsfig{file=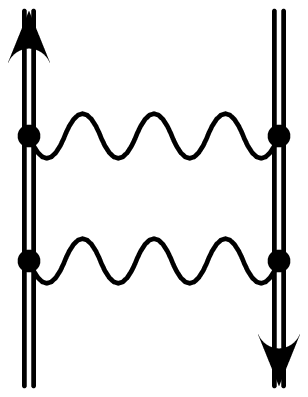,width=1cm}\end{minipage}}
+\cdots=\exp\left(\mbox{\begin{minipage}{1.05cm}
\epsfig{file=qed1.ps,width=1cm}\end{minipage}}\right):
\end{equation}
multi-photon exchanges exponentiate!
The solid lines denote the static sources and an integration of the
vertices over all positions along the contour of the Wilson loop
is understood to be included within the notation.
The static potential in the limit of large Euclidean times
$\tau$ is related to the
Wilson loop,
\begin{equation}
\label{eq:wil}
\langle W(r,\tau)\rangle
\propto\exp[-V_0(r)\tau]\qquad(\tau\rightarrow\infty),
\end{equation}
which means that only contributions whose
logarithms are proportional to $\tau$ have to be considered;
interactions with the spatial closures of the loop can be ignored.
The single photon exchange within
the exponent on the right hand side of Eq.~(\ref{eq:pert})
can easily be calculated: a factor $\tau$ is obtained by translational
invariance and the remaining time integral yields a $\delta$-function
for the 4-component of the photon momentum. After a 3-dimensional
Fourier transform one finds the well known result,
$V_0=-\alpha_{\mbox{\scriptsize fs}}/r$ to all orders
of perturbation theory.

The situation becomes somewhat more involved if fermions are included,
due to the occurrence of diagrams such as,
\begin{equation}
D=\mbox{\begin{minipage}{1.55cm}
\epsfig{file=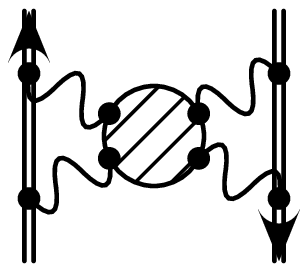,width=1.5cm}\end{minipage}}.
\end{equation}
However, irreducible diagrams $D_i$ still exponentiate and
the relation, 
$V_0(r)=-\lim_{\tau\rightarrow\infty}\sum_idD_i(r,\tau)/d\tau$, holds.
In a non-Abelian theory, starting from a two gluon
exchange, the colour pre-factor of a
given diagram depends on the ordering of the vertices along the
contour of the Wilson loop. Hence,
contributions are encountered that do not exponentiate
anymore.~\cite{Fischler:1977yf}

Still, at least up to order $\alpha_s^3$,
the static potential $V_0$ agrees with the result one obtains ignoring
gluons that couple to the spatial closures of the Wilson loop,
the so-called {\em singlet potential} $V_s$.
Within perturbation theory
it is also possible to define a so-called {\em octet potential},
$V_o$, by introducing colour generators at zero and infinite times into
the colour traces. In doing so, one finds
$V_o=-V_s/8$, at least to order $\alpha_s^2$. It is possible to calculate
hybrid potentials in perturbation theory too, starting from
linear combinations of Wilson loops where the spatial connections are deformed
or colour fields are inserted. While for instance
the hybrid potential $V_{\Pi_u}$ vanishes to order $\alpha_s$,
the next excited state, $V_{\Sigma_u^-}$ agrees with $V_o$ to this order.

Up to order $\alpha_s^3$ non-Abelian contributions
do not cause any fundamental
problem, however, it has early been noticed~\cite{Appelquist:1977tw}
that diagrams,
\begin{equation}
D_{\mbox{\scriptsize ADM}}=\mbox{\begin{minipage}{2.05cm}
\epsfig{file=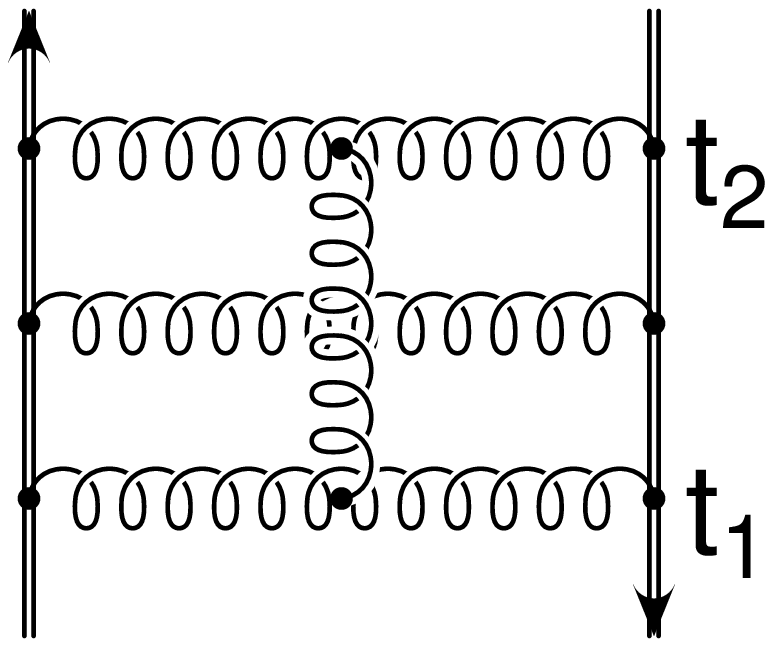,width=2cm}\end{minipage}}
\propto N^2C_D^2\alpha_s^4\tau\int_0^\tau\!\!\!\!
dt\,\frac{1}{r^2+t^2}\frac{t}{r}
\propto\alpha_s^4\tau\ln\tau,
\end{equation}
diverge faster than $\tau$ as $\tau\rightarrow\infty$.
This means that a na\"{\i}ve perturbative expansion of the
static potential encounters an infra red problem:
between the times $t_1$ and $t_2$ the quark and anti-quark
are in a colour octet state and the
first argument of the time integral should be
proportional to $\exp(-\Delta Vt)$ with $\Delta V\simeq V_o-V_s$,
which is small for times $t=t_2-t_1\gg\Delta V^{-1}$.
However, in perturbation theory the real space propagator
for large times only decays like $1/(r^2+t^2)$ and this
(wrong) behaviour finally causes the infra red divergence. 
It has therefore been suggested
to regulate the encountered divergence by resumming
all possible
interactions~\cite{Appelquist:1977tw} between $t_1$ and $t_2$.

Since the infra red problem is related to the emission of
an ultra-soft gluon, it has also been equated to the Lamb shift
in the literature.~\cite{Thacker:1991bm}
One should, however, keep in mind that
in QED the Lamb shift affects atomic bound states and not the
Coulomb potential itself (which in QCD happens to be a bound state too).
We shall see below that
another ``Lamb shift'', analogous to that of QED exists within
quarkonia systems. One should also remember that while
perturbation theory generically encounters problems when
applied to bound states, the static potential is perfectly well defined
in a non-perturbative context.
Despite all problems with the static potential itself,
Potential Nonrelativistic QCD~\cite{Brambilla:2000xf} (pNRQCD)
might allow for a consistent and systematic perturbative
treatment of heavy quark bound states.
Within the pNRQCD Lagrangian
a {\em heavy quark} singlet potential
$V_s^{\mbox{\scriptsize pNRQCD}}$ appears that can be
related to the {\em static} singlet potential $V_s$. It has been
demonstrated~\cite{Brambilla:1999qa} that the divergence encountered in $V_s$
at order
$\alpha_s^4$ is cancelled by a counter term from the matching
of pNRQCD to QCD: the resulting
$V_s^{\mbox{\scriptsize pNRQCD}}$ does not incorporate any
singlet-octet-singlet
transitions which are dealt with explicitely in the
pNRQCD multipole expansion.

Finally, I mention the
remarkable fact that up to order~\cite{Schroder:1999vy} $\alpha_s^3$
potentials between charges within different representations $D$
of the gauge group scale in proportion to the respective
Casimir factors, $C_D=(1/D)\mbox{tr}\,T_a^DT_a^D$.
This Casimir scaling seems to be rather accurately satisfied
in the non-perturbative regime too.~\cite{Bali:2000un}
\section{The ``heavy quark potential'' from QCD}
\label{sec:heavy}
The natural starting point for deriving a non-relativistic Hamiltonian
for quark\-onia bound states is an effective field theory,
NRQCD:~\cite{Caswell:1986ui,Thacker:1991bm}
\begin{equation}
\label{eq:nrqcd}
{\mathcal L}_{\mbox{\scriptsize NRQCD}}=
-\psi^{\dagger}\left[m_Q+D_4-c_2\frac{{\mathbf D}^2}{2m_Q}+
\cdots\right]\psi+\chi^{\dagger}
\left[\cdots\right]\chi+{\mathcal L}_{\mbox{\scriptsize glue}}.
\end{equation}
$\psi$ and $\chi$ are the quark and antiquark Pauli spinors.
We eliminate the matching constant
$c_2$ by defining $m=m_Q/c_2=m_Q+\delta m$, where
$m_Q$ is the bare quark mass appearing in the original Dirac
Lagrangian in Euclidean space,
${\mathcal L}= \bar{q}(D\!\!\!\!/+m_Q)q$.
The underlying idea of an effective field theory description
is that if the quark mass is much bigger than
exchange momenta and binding energy,
QCD processes of scales $q\geq \mu>mv$ can be
integrated out into local matching constants $c_i$.

For systems containing only one heavy quark the leading order
effective Lagrangian will not incorporate the kinetic term
and the usual heavy quark effective theory (HQET) power counting
in powers of the inverse quark mass applies.
However,
heavyonia bound states cannot be addressed
within a purely static theory, basically because, without a
kinetic term, quarks
that travel at different velocities will never interact with each other:
hence, the leading order Lagrangian is that of Eq.~(\ref{eq:nrqcd}).
Since the static theory does not apply to quarkonia systems
it is also not {\em a priori} clear whether the
phenomenological potential $V_{\mbox{\scriptsize phen}}$ of
Eq.~(\ref{eq:schroe})
is at all related to the static
potential $V_0$ in some approximation.~\cite{Brown:1979ya}
To keep the discussion brief I will use the conventional
power counting rules~\cite{Thacker:1991bm}
in the relative heavy quark velocity $v$
throughout this article.
Several well founded objections to this scheme
exist and other counting schemes
have been suggested to be more suitable or better
defined.~\cite{Grinstein:2000xb,Bali:2000gf,Luke:2000kz,Brambilla:2000xf,antonio}
According to the ``standard'' conventions the leading order NRQCD Lagrangian
is counted as order $v^2$.

In Table~\ref{tab:scales} estimates of various scales for
the charmonium, bottomonium and (unstable and, therefore, hypothetical)
toponium ground states from
a potential calculation~\cite{Bali:2000gf}
are displayed. For comparison, we include
the corresponding estimates for positronium.
While binding energies and level splittings are of size $mv^2$
(ultra-soft), the momenta exchanged between the quarks are of
order $mv$ (soft). NRQCD is the effective theory for physics at scales
$q<\mu$, $mv\leq\mu\leq m$, while physics of scales $q>\mu$,
has been integrated
out into
matching coefficients $c_i(m/\mu,\alpha_s)$
that are in principle calculable from the QCD Lagrangian.
It is possible to integrate out the soft scale $mv$ in a further step with the
result of the effective theories, pNRQCD~\cite{Brambilla:2000xf}
or vNRQCD.~\cite{Luke:2000kz}

\begin{table}[t]
\caption{Hard, soft and ultra-soft scale estimates.~\protect\cite{Bali:2000gf}
\label{tab:scales}}
\vspace{0.2cm}
\begin{center}
\begin{tabular}{|c|c|c|c|c|}
\hline
&$J/\psi$&$\Upsilon$&$t\bar{t}$&$e^+e^-$\\\hline
$m$&1.4~GeV&4.7~GeV&175~GeV&511~keV\\
$mv$&0.7~GeV&1.3~GeV&45~GeV&3.7~keV\\
$mv^2$&0.4~GeV&0.4~GeV&12~GeV&0.027~keV\\
$v$&0.5&0.29&0.26&0.007\\\hline
\end{tabular}
\vskip -.5cm
\end{center}
\end{table}

In NRQCD, being a nonrelativistic theory, the time evolution of
a quark in a background gauge field, $\{A_{\mu}(x)\}$,
is controlled by a Hamiltonian,
\begin{equation}
H_{\psi}=m-\delta m+igA_4-\frac{{\mathbf D}^2}{2m}+\cdots.
\end{equation}
The quark propagator, $K=\psi\psi^{\dagger}$, obeys the evolution equation,
$-d_4K=H_{\psi}K$, which for the initial condition,
$\left.K(x,y)\right|_{x_4=y_4}=\delta^3({\mathbf x}-{\mathbf y})$,
is solved by,
\begin{equation}
\label{eq:propa}
K(x,y)\propto
\int_{{\mathbf z}(y_4)={\mathbf y}}^{
{\mathbf z}(x_4)={\mathbf x}}\!\!\!\!\!
D{\mathbf z}\,
D{\mathbf p}\,\exp\left\{
\int_{y_4}^{x_4}\!\!dt\,\left[
{\mathbf p}\dot{{\mathbf z}}-
H_{\psi}({\mathbf z},{\mathbf p})\right]
\right\},
\end{equation}
where the path integral is over all paths connecting $y$ with
$x$.

For illustrative purposes we sketch the derivation of
a quantum mechanical Hamiltonian for quarkonium systems
to leading order in $v$:
we combine two propagators for quark and antiquark into a
four-point function,\\[-.3cm]
\begin{equation}
\label{eq:green}
G({\mathbf r},\tau;\cdots)
=\mbox{\begin{minipage}{3.1cm}\epsfig{file=4point.eps,width=3cm}
\end{minipage}}
=\frac{1}{3}\mbox{Re}\,\mbox{tr}\left\{U_0^{\dagger}
K(y_1,x_1)
U_{\tau}K(x_2,y_2)\right\}
\end{equation}\\[-.3cm]
$U_{\tau}$ denotes a gauge transporter at time $\tau$.
To higher orders in the velocity expansion
$G$ will not only depend on the distance
but also on spin, angular momentum and momentum.
It can be shown~\cite{Barchielli:1990zp} that to leading order only 
the shortest paths between $y_i$ and $x_i$ contribute to the
path integral Eq.~(\ref{eq:propa}), such that
after factorising the gauge field dependent and independent parts one obtains,
\begin{equation}
\langle G({\mathbf r},\tau)\rangle
\propto\exp\left[-\int_0^{\tau}\!\!\!dt\,\sum_{j=1}^2
\left(m_j-\delta m_j+\frac{{\mathbf p}_j^2}{2m_j}\right)\right]\langle
W({\mathbf r},\tau)\rangle,
\end{equation}
where $W({\mathbf r},\tau)$
is the usual Wilson loop  and  the expectation value
is over gauge field configurations.
Now the Hamiltonian $H$ that governs the evolution of
the two-particle state, $H\langle G\rangle = -d_4\langle G\rangle$
can easily be read off,
\begin{equation}
\label{eq:hami}
H=2[m-\delta m(\mu)]+\frac{{\mathbf p}^2}{2\mu_R}+V_0(\mu;r),
\end{equation}
where we have set $m=m_1=m_2=2\mu_R$. Indeed, the leading order
Hamiltonian is that of
Eq.~(\ref{eq:schroe}) with
$V_{\mbox{\scriptsize phen}}(r)=V_0(\mu;r)-2\delta m(\mu)$.
The quantum mechanical Hamiltonian should not depend
on the scale $\mu$. This implies the relation,
$
2\delta m(\mu')=
2\delta m(\mu)+
V_{\mbox{\scriptsize self}}(\mu')-V_{\mbox{\scriptsize self}}(\mu)$,
between the mass shift $\delta m(\mu)$ between ``kinetic''
mass $m$ and QCD quark mass $m_Q(\mu)$ and the static self energy
$V_{\mbox{\scriptsize self}}(\mu)$, defined in Eq.~(\ref{eq:vself}).
\section{The ``Lamb shift'' and relativistic corrections}
The assumption, $\Delta t_g > \Delta t_Q$, underlying the
adiabatic approximation, is clearly violated whenever ultra-soft
gluons with momenta smaller than the interaction energy
are exchanged; at the end of an interaction time of order
$\Delta t_Q$ such a gluon will still be present and cannot be integrated
out into a potential. In the positronium case
such effects result in the Lamb shift which can be accounted for
by considering transitions between
higher Fock states, $|e^+e^-\gamma\rangle,
|e^+e^-\gamma\gamma\rangle, \ldots$
and $|e^+e^-\rangle$.
Any formalism that does not incorporate such transitions
can only be approximate.
Where then is the loop hole of the derivation
of Sec.~\ref{sec:heavy} above?
The definition of the 4-point Green function
Eq.~(\ref{eq:green}) is ambiguous since the
$D_{\infty h}$ states of the gluonic connections $U_{0}$ and $U_{\tau}$
are unspecified. In general,
$G$ will carry two more indices that run over gluonic excitations,
$a,b=\Sigma_g^+,\Pi_u,\Sigma_u^-,\cdots$, in addition to
time, distance, spin, momentum and angular momentum of the quarks.
The derivation of Eq.~(\ref{eq:hami}) can then be generalised,
resulting in the Schr\"odinger equation~\cite{Bali:2000gf}
$H_{ab}\psi^b_n=E\psi^a_n$, where the $J^{PC}$ quantum numbers of
quarkonia eigenstates $\psi_n$ are products of quark and
gluon quantum numbers. Physical states will be linear
combinations of standard quark model states ($\Sigma_g^+$)
and various hybrid excitations.

\begin{table}
\caption{Combinations of spins and angular momenta that can couple
to $J^{PC}=1^{--}$.\label{tab:jpc}}\vspace{.2cm}
\begin{center}
\begin{tabular}{|c|cccccccc|}
\hline
$\Lambda^{\sigma_v}_{\eta}$&${\mathbf \Sigma_g^+}$
&$\Sigma_u^-$
&$\Pi_u$
&$\Sigma_g^-$
&$\Pi_u$
&${\mathbf \Sigma_g^+}$
&$\Pi_u$
&$\Delta_g$\\\hline
$K$&$S$&$P$&$P$&$P$&$P$&$D$&$D$&$D$\\\hline
$S$&$1$&$0$&$0$&$1$&$1$&$1$&$1$&$1$\\\hline
\end{tabular}
\vskip -.5cm
\end{center}
\end{table}

The off-diagonal elements $H_{\Sigma_g^+b}$ mediate transitions
between standard states and states with excited
glue and are related to gluon emission by the valence quarks.
Each such element is accompanied by a pre-factor $gv$, $g=\sqrt{4\pi\alpha_s}$.
We may therefore treat mixing effects as perturbations.
Let us start from the unperturbed, diagonal Hamiltonian,
\begin{equation}
H_{aa}=2(m-\delta m)+\frac{{\mathbf p}_a^2}{2\mu_R}+V_a(r),
\quad H_{aa}\psi_n^{0,a}=E_n^{0,a}\psi_n^{0,a}.
\end{equation}
The energy shift of the state $n$ of the
lowest lying channel ($a=\Sigma_g^+$) is,
\newpage
\begin{equation}
\label{eq:pertu}
\Delta E_n^{a}=\sum_{m,b\neq a}
\frac{\left|\left\langle\psi_n^{0,a}\left|H_{ab}\right|\psi_m^{0,b}
\right\rangle\right|}{E_m^{0,b}-E_n^{0,a}}=
{\mathcal O}\left(\alpha_sv^2E_n^{0,a}\right),
\end{equation}
where only states with equal $J^{PC}$ content yield a non-vanishing
matrix element in the nominator
and $E_n=E_n^{0,\Sigma_g^+}+\Delta E_n^{\Sigma_g^+}+{\mathcal O}(v^6)$.
In Table~\ref{tab:jpc} we have listed
all possible combinations of gluonic $D_{\infty h}$ excitation
$\Lambda^{\sigma_v}_{\eta}$, total angular momentum $K$ and spin $S$
of a vector particle ($J^{PC}=1^{--}$). It turns out that to order $v^4$
only mixing of the standard $S$ wave state with the
$\Sigma_u^-$ $P$ wave state has to be considered.~\cite{Bali:2000gf}
In this formalism spin exotic states like $1^{-+}$
that do not have any $\Sigma_g^+$
content are automatically accounted for.

Having understood these mixing effects,
we can now say that ``valence gluons'' that accompany the quarks
in the form of hybrid excitations of the flux tube and ``sea gluons'',
whose average effect can be integrated out into
an interaction potential, can be distinguished from each other. To
lowest order of the relativistic
expansion pure quark model quarkonia and pure quark-gluon hybrids exist,
which then undergo mixing with each other as higher orders
in $v$ are incorporated.
Moreover, at leading order the Hamiltonian
is of Schr\"odinger type and $V_{\mbox{\scriptsize phen}}=V_0$.
As soon as one allows for higher orders in the velocity expansion,
an instantaneous interaction potential does not exist anymore
but the corrections can systematically be accounted for.

With respect to QED the Lamb shift is na\"{\i}vely enhanced
by a relative factor, $\alpha_sv^2_{\Upsilon}/(\alpha_{\mbox{\scriptsize fs}}
v^2_{e^+e^-})\approx 6\times 10^4$.
However, below a certain glueball (or, when sea quarks are included,
meson) radiation threshold, the spectrum of
excitations is (unlike in the QCD case) discrete and the denominator
of Eq.~(\ref{eq:pertu}) will become large.
It is clear that
as long as gaps between hybrid potentials are bigger than
quarkonia level splittings with respect to the $1S$ state,
transitions will be energetically penalised: the wave function
will only see the lowest lying potential and the adiabatic
approximation is reliable while one might expect
levels such as $\Upsilon(3S)$ and $\Upsilon(4S)$ to be affected
by the presence of hybrid excitations.

We notice formal similarities between the discussion above
and the pNRQCD Lagrangian that includes transition
elements between two different states, singlet and octet.
However, to draw the analogy, singlet~$\simeq\Sigma_g^+$ and
octet~$\simeq$~hybrid, is not straight forward.
Moreover, in the potential framework transitions between
different $D_{\infty h}$ states are only possible via interactions
between the valence quarks and the glue while in pNRQCD
an ultra-soft gluon that causes a singlet-octet transition
can also be radiated from another gluon.
Effects of this latter kind are already
included automatically in the non-perturbative matrix
elements $H_{ab}$, defined through expectation values of Wilson loop like
operators.

The velocity
will approach zero only as a
logarithmic function of the quark mass: 
the numbers of
Table~\ref{tab:scales} reveal that
$v^2$ is reduced from 0.085 to just 0.070 when
replacing bottom quarks by (almost forty times heavier) top quarks.
Therefore, although $v\rightarrow 0$ as $m\rightarrow\infty$,
at any reasonable quark mass value one has to address
Lamb shift effects as well as standard relativistic
corrections which are also mandatory for
an understanding of fine structure splittings.

The complete Hamiltonian to order $v^4$ is
known~\cite{Eichten:1979pu,Barchielli:1990zp,Bali:1997am,Bali:2000gf,Pineda:2000sz,antonio}
where various
corrections are given in
terms of expectation values of Wilson loop like operators,
most of which have been calculated on the
lattice.~\cite{Bali:1997am,Bali:2000gf}
In addition to Pauli, Thomas and Darwin like terms, momentum dependent
corrections appear due to insertions of the operators
$D_4$ and ${\mathbf D}^2/(2m)$ as well as $1/m$ and $1/m^2$
corrections that are spin- and momentum-independent.~\cite{Pineda:2000sz}
We remark that the momentum dependent corrections have nothing to do with
the Lamb shift.
The derivation can either be performed along the lines of
Sec.~\ref{sec:heavy}~\cite{Brown:1979ya,Eichten:1979pu,Barchielli:1990zp}
or in terms of quantum mechanical
perturbations.~\cite{Bali:2000gf,Pineda:2000sz,antonio}
The predictive power is at present
in particular limited by an insufficient
knowledge of the matching constants between QCD and NRQCD.
For instance we estimate~\cite{Bali:1999pi} a related uncertainty of 25~\%
on $\Upsilon$ fine structure splittings 
while relativistic order $v^6$ corrections are unlikely
to change order $v^4$ QCD potential predictions by more than 10~\%.
\section{Summary}
We have demonstrated that in
a non-relativistic situation, it is possible to factorise
gluonic effects from the slower dynamics of the quarks
and to derive ``potential models'' from QCD.
Effective field theory methods
turn out to be essential for this step.
The resulting Hamiltonian representation of the bound state problem
in terms of functions of canonical variables
offers a very intuitive and transparent representation of
quarkonium physics. 
It highlights parallels as well as differences
to well understood atomic physics. 

The adiabatic approximation is violated when ultra-soft gluons are
radiated, i.e.\ when the nature of the bound state changes during
the interaction time. Such effects can systematically be incorporated
into the potential formulation by enlarging the basis of states
onto which the Hamiltonian acts
and a lattice study of the relevant transition matrix elements
$H_{ab}$ is on its way.
The validity of the
adiabatic approximation is tied to that of the non-relativistic
approximation in so far as off-diagonal entries of
$H_{ab}$ are suppressed by powers of the velocity, $v$.
Hybrid mesons become a well defined concept in the potential
approach and translation into the variables
used for instance in flux tube models is straight forward.
In view of phenomenological applications,
a precise determination
of the matching coefficients between QCD and (lattice) NRQCD is
required.
\section*{Acknowledgements}
I thank the organisers of the conference for the invitation.
I have been supported by EU grant HPMF-CT-1999-00353.

\end{document}